\documentclass[prc,twocolumn,showpacs,preprintnumbers,amsmath,amssymb]{revtex4-1}
\usepackage[dvips]{graphicx}
\usepackage{dcolumn}
\usepackage{bm}
\begin{document}
\title{Effects of ground-state correlations on magnetic dipole excitations in $^{40}$Ca}
\author{Mitsuru Tohyama}
\affiliation{Faculty of Medicine, Kyorin University, Mitaka, Tokyo
  181-8611, Japan     }
\begin{abstract}
The effects of ground-state correlations on the magnetic dipole excitations in $^{40}$Ca are studied using an extended
random phase approximation (ERPA) derived from the time-dependent density-matrix theory. Comparison is made with other extended RPA approaches, the renormalized RPA, the self-consistent RPA
and the extended second RPA which also include the effects of ground-state correlations. It is pointed out that direct
excitations from two particle - two hole space which 
are properly treated in ERPA cause strong magnetic dipole transitions in $^{40}$Ca.
\end{abstract}
\maketitle
\section{Introduction}
Strong magnetic dipole transitions have been observed in doubly $LS$ closed-shell nuclei such as $^{16}$O \cite{snov,kuch} and $^{40}$Ca \cite{gross,burt} and the importance of 
ground-state correlations (core excitations) has been pointed out \cite{alex} since the magnetic dipole transitions are not allowed under 
the Hartree-Fock (HF) assumption for the ground states of these nuclei. 
The random phase approximation (RPA) cannot deal with the magnetic dipole transitions in these nuclei because it is based on the HF ground state.
Various attempts have been made to incorporate the effects of ground-state correlations into RPA, where
the effects of ground-state correlations are expressed by 
the fractional occupation probability $n_\alpha$ of a single-particle state $\alpha$ and the correlated part $C_2$ of a two-body density matrix.
The renormalized RPA (rRPA) \cite{rowe1,rowe2} includes $n_\alpha$, which opens new particle (p)--p and hole (h)--h transitions.
The self-consistent RPA (SCRPA) \cite{scrpa1,scrpa2} includes both $n_\alpha$ and $C_2$: $C_2$ plays a role in modifying
the energies of p--h, h--h and p--p pairs and also the p-h, p--p and h--h correlations. The coupling to a two-body amplitude is not taken into account in rRPA and SCRPA, however.
The response function formalism of Refs. \cite{taka1,taka2} uses 
perturbatively calculated $n_\alpha$ and $C_2$ and includes the coupling to the two-body amplitude.
The extended second RPA (ESRPA) of Refs. \cite{nishi,srpa} implements $n_\alpha$ and $C_2$ in the second RPA equation in a way similar to the response function formalism.
We have developed an extended RPA (ERPA) from the small amplitude limit of the time-dependent density-matrix theory (TDDM) \cite{WC,GT}.
ERPA consists of the coupled equations for the one-body and two-body transition amplitudes and
includes both $n_\alpha$ ad $C_2$ as the ground-state correlation effects.
ERPA has been applied to electric dipole and quadrupole excitations in oxygen and calcium isotopes \cite{toh07,toh17} and it has been shown that ground-state correlations play an important role in
enhancing the fragmentation of dipole and quadrupole strengths. 
In this paper we extend the application of ERPA to the magnetic dipole transitions in $^{40}$Ca. Comparing ERPA with rRPA, SCRPA and ESRPA, we 
clarify important ingredients to be included in the extension of RPA to treat the magnetic dipole excitations.
The paper is organized as follows. The formulation of ERPA is presented in sect. 2,
numerical details are explained in sect. 3, the obtained results are given in sect. 4 and sect. 5 is devoted to summary.

\section{Formulation}
The ground state used in ERPA is given as a stationary solution of the TDDM equations. 
The TDDM equations consist of the coupled equations of motion for the one-body density matrix $n_{\alpha\alpha'}$ 
(the occupation matrix) and the correlated part of the two-body density matrix $C_{\alpha\beta\alpha'\beta'}$
($C_2$). The equations of motion for reduced density matrices form
a chain of coupled equations known as the Bogoliubov-Born-Green-Kirkwood-Yvon (BBGKY) hierarchy. In TDDM the BBGKY
hierarchy is truncated by replacing a three-body density matrix with anti-symmetrized products of the one-body and
two-body density matrices \cite{WC,GT}. The TDDM equation for $C_{\alpha\beta\alpha'\beta'}$ contains all effects of two-body correlations;
p-p, h-h and p-h correlations. 
A stationary solution of the TDDM equations can be obtained by using the gradient method \cite{toh07}.

The ERPA equations are derived as the small amplitude limit of TDDM and are written in matrix form
for the one-body and two-body amplitudes $x^\mu_{\alpha\alpha'}$ and $X^\mu_{\alpha\beta\alpha'\beta'}$ \cite{toh07}
\begin{eqnarray}
\left(
\begin{array}{cc}
A&B\\
C&D
\end{array}
\right)\left(
\begin{array}{c}
{x}^\mu\\
{X}^\mu
\end{array}
\right)
=\omega_\mu
\left(
\begin{array}{cc}
S_{1}&T_{1}\\
T_{2}&S_{2}
\end{array}
\right)
\left(
\begin{array}{c}
{x}^\mu\\
{X}^\mu
\end{array}
\right),
\label{ERPA1}
\end{eqnarray}
where $A$, $B$, $C$ and $D$ are given as the expectation values of the double commutators between Hamiltonian and either one-body or two-body excitation operators while
$S_1$, $T_1~(=T_2^\dag)$ and $S_2$ are the expectation values of the commutators between either one-body or two-body excitation operators. The effects of ground-state correlations are included
in Eq. (\ref{ERPA1}) through $n_\alpha$ and $C_2$.
Each matrix element of Eq. (\ref{ERPA1}) is given explicitly in Ref. \cite{ts08}.
The orth-normalization is given by
\begin{eqnarray}
\left({x}^{\mu *}~~
{X}^{\mu *}
\right)
\left(
\begin{array}{cc}
S_{1}&T_{1}\\
T_{2}&S_{2}
\end{array}
\right)
\left(
\begin{array}{c}
{x}^\nu\\
{X}^\nu
\end{array}
\right)=\pm\delta _{\mu\nu},
\label{norm}
\end{eqnarray}
where $({x}^{\mu *}~~{X}^{\mu *})$ is the left eigenvector of Eq. (\ref{ERPA1}) and the minus sign is for negative-energy states.
The one-body sector of Eq. (\ref{ERPA1}) $A{x}^\mu=\omega_\mu S_1{x}^\mu$ is the same as the equation in SCRPA and the neglect of $C_2$ in the SCRPA
equation reduces to the rRPA equation.
If the HF assumption is made for the ground state,  Eq.(\ref{ERPA1}) is  reduced to
the SRPA equation \cite{srpa} though it is irrelevant to the magnetic dipole excitations in $^{40}$Ca.
When all $C_2$'s except for those in $A$ on the left-hand side of Eq. (\ref{ERPA1}) are neglected, 
Eq. (\ref{ERPA1}) corresponds to the ESRPA equation of Refs. \cite{nishi,srpa}. The interaction in $D$ is also neglected in ESRPA.

\section{Calculational details}
The occupation probability $n_{\alpha}$ and $C_{2}$ are
calculated within TDDM using a truncated single-particle basis:
The $2s_{1/2}$, $1d_{3/2}$, $1d_{5/2}$ and $1f_{7/2}$ states are used for both protons and neutrons.
For the calculations of the single-particle states we use the Skyrme III force \cite{skIII}.
To reduce the dimension size, we only consider the 2p--2h and 2h--2p elements of $C_{2}$.
A simplified interaction which contains only the $t_0$ and $t_3$ terms of the Skyrme III force is used as the residual interaction \cite{toh07}.
The spin-orbit force and Coulomb interaction are also omitted from the residual interaction.
The magnetic dipole states are obtained by using Eq. (\ref{ERPA1}).
The one-body amplitudes ${x}^\mu_{\alpha\alpha'}$ are defined with a large number of single-particle states including those in the 
continuum: We discretize the continuum states by confining the wavefunctions in a sphere with radius 15 fm and take all 
the single-particle states with $\epsilon_\alpha\le 50$ MeV and 
$j_\alpha\le 9/2 \hbar$. 
As the residual interaction in Eq. (\ref{ERPA1}), we use a force of the same form as that used in the ground-state
calculation.
Since the residual interaction is not consistent with the effective interaction used in the calculation of the single-particle states,
it is necessary to reduce the strength of the residual interaction so that the spurious mode corresponding
to the center-of-mass motion comes at zero excitation energy in RPA. We found that
the reduction factor $f$ is 0.66. The residual interaction used in the matrices $A$, $B$ and $C$ which involve the couping to the one-body amplitudes is multiplied with this factor $f$.
We include all p-h and h-p amplitudes. For p--p and h--h we include the amplitudes with $|n_{\alpha}-n_{\alpha'}|\ge 0.05$. 
For the single-particle states used in the ground-state calculation we also include the diagonal part ${x}^\mu_{\alpha\alpha}$, which plays an important role in increasing transition strength as will be
shown below.
To reduce the number of the two-body amplitudes, we consider only  
the 2p--2h and 2h--2p components of ${X}^\mu_{\alpha\beta\alpha'\beta'}$ using
the $2s_{1/2}$, $1d_{3/2}$, $1d_{5/2}$, $1f_{7/2}$ $1f_{5/2}$, $2p_{3/2}$ and $2p_{1/2}$ states for both protons and neutrons.

The full M1 excitation operator is the following.
\begin{eqnarray}
\hat{\cal Q}({\rm M1})=\left(\frac{3}{4\pi}\right)^{\frac{1}{2}}\sum_i[\alpha_l{\bm l}_i+\alpha_\sigma{\bm \sigma}_i+(\beta_l{\bm l}_i+\beta_\sigma{\bm \sigma}_i)\tau_i^z],
\nonumber \\
\end{eqnarray}
where ${\bm l}_i$, $\frac{1}{2}{\bm \sigma}_i$ and $\frac{1}{2}{\bm \tau_i}$ are the nucleon orbital angular momentum, spin and isospin, respectively.
The coefficients are give by $\alpha_l=-\beta_l=\mu_0/2$, $\alpha_\sigma=(g_{\rm n}+g_{\rm p})/4$ and $\beta_\sigma=(g_{\rm n}-g_{\rm p})/4$
with the nucleon spin $g$-factors $g_{\rm n}=-3.826\mu_0$ and $g_{\rm p}=5.585\mu_0$. Here $\mu_0$ is the nuclear magneton. 
In the following analysis we use the dominant term given by ${\bm \sigma}_i\tau_i^z$ \cite{adachi,sagawa}:
In our calculations shown below this operator explains about 82 \% 
of the transition strength of isovector magnetic dipole modes. 

\section{Results}
\subsection{Ground state}
\begin{table}
\caption{Single-particle energies $\epsilon_\alpha$ and occupation probabilities 
$n_{\alpha\alpha}$ calculated in TDDM for $^{40}$Ca.}
\begin{center}
\begin{tabular}{c rr rr} \hline
 &\multicolumn{2}{c}{$\epsilon_\alpha$ [MeV]}&\multicolumn{2}{c}{$n_{\alpha}$}\\ \hline 
orbit & proton & neutron  & proton & neutron  \\ \hline
$1d_{5/2}$ & -15.6 & -22.9 & 0.923 & 0.924  \\
$1d_{3/2}$ & -9.4 & -16.5 & 0.884 & 0.884  \\
$2s_{2/2}$ & -8.5 & -15.9 & 0.846 & 0.846   \\ 
$1f_{7/2}$ & -3.4 & -10.4 & 0.154 & 0.154  \\ 
$1f_{5/2}$ &5.2 & -1.3 & -& - \\\hline
\end{tabular}
\label{tab1}
\end{center}
\end{table}

The occupation probabilities calculated in TDDM for $^{40}$Ca are shown in Table \ref{tab1}.
The deviation from the HF values ($n_{\alpha\alpha}$=1 or 0) is more than 10\%,
which means that the ground state of $^{40}$Ca is a strongly correlated state as was pointed out in an RPA approach \cite{agassi}
and perturbation calculations \cite{srpa,adachi}.

\begin{figure} 
\begin{center} 
\includegraphics[height=6cm]{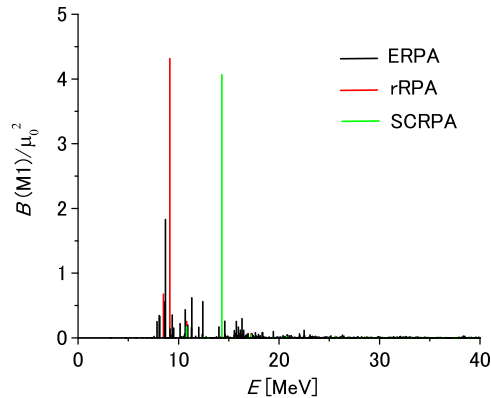}
\end{center}
\caption{M1 strength distribution calculated in ERPA (black lines), rRPA (red lines) and SCRPA (greenlines) for $^{40}$Ca.} 
\label{cam1} 
\end{figure} 
\begin{figure} 
\begin{center} 
\includegraphics[height=6cm]{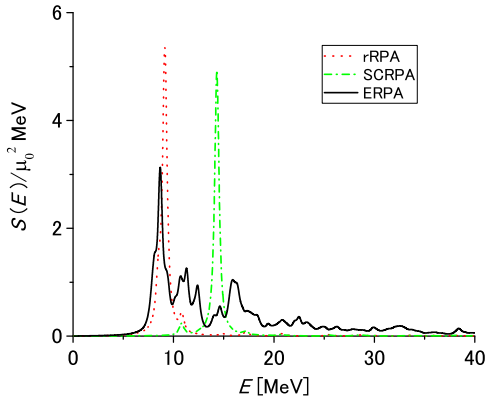}
\end{center}
\caption{Strength functions calculated in ERPA (solid line), rRPA (dotted line) and SCRPA (dot-dashed line) for the magnetic dipole excitation in $^{40}$Ca. 
The distributions are smoothed with an artificial width $\Gamma=0.5$ MeV.} 
\label{cam2} 
\end{figure} 

\subsection{Magnetic dipole excitation}
The M1 strengths ($B({\rm M}1)\uparrow$) calculated in ERPA (black lines), rRPA (red lines) and SCRPA (green lines) are shown in Fig. \ref{cam1}.
To facilitate comparison, we also display in Fig. \ref{cam2} the distributions smoothed with an artificial width $\Gamma=0.5$ MeV.
The occupation probability $n_{\alpha}$ used in rRPA, $n_{\alpha}$ and $C_2$ used in SCRPA are not self-consistently determined by the one-body amplitudes but are taken from the results of the TDDM ground-state calculation.
First we discuss the results in rRPA and SCRPA.
There are two states below 10 MeV in rRPA. The state at 9.1 MeV has 4.32 $\mu_0^2$ and mainly consists of the neutron $1f_{7/2}\rightarrow 1f_{5/2}$ component: The contribution of this component to the normalization
Eq. (\ref{norm}) is $84.9~\%$ and that of the proton $1f_{7/2}\rightarrow 1f_{5/2}$ component $14.9~\%$.
The state at 8.5 MeV has 0.67 $\mu_0^2$ and consists of $85.0 \%$ of the proton $1f_{7/2}\rightarrow 1f_{5/2}$ component and $14.9 \%$ of the neutron $1f_{7/2}\rightarrow 1f_{5/2}$ component
as measured by the contributions to the normalization. Thus the proton and neutron transitions are separated due to the asymmetry in the single-particle energy. 
Note that the proton $1f_{5/2}$ state is in the continuum (see Table \ref{tab1}).
SCRPA gives the collective state at 14.3 MeV with 4.07 $\mu_0^2$. The contributions of the proton and neutron $1f_{7/2}\rightarrow 1f_{5/2}$ transitions 
to the normalization Eq. (\ref{norm}) are 38.5 \% and 59.7 \%, respectively.
Thus, in rRPA and SCRPA the M1 states with the largest transition strength consist of pure $1f_{7/2}\rightarrow 1f_{5/2}$ transitions. In the case of SCRPA the M1 state gains extra excitation energy
due to the self-energy contributions. As has been pointed out in Ref. \cite{toh13}, 
the coupling to the
two-body amplitudes is needed to decrease the excitation energy of the collective state when both effects of $n_\alpha$ and $C_2$ are included in the one-body sector of Eq. (\ref{ERPA1}).
Experimentally, two M1 states with $T=1$ have been observed at 9.87 MeV and 10.32 MeV with $B({\rm M}1)\uparrow=0.23\pm0.06~\mu_0^2$ 
and $1.17\pm0.06~\mu_0^2$, respectively \cite{gross,burt,allen,pringle}. To obtain better agreement with the observed excitation energies in the framework of rRPA (and ERPA), 
we need to adjust the 
spin-orbit force and (or) include other spin-dependent interactions which are omitted in Skyrme III. We found that
the spin-isospin dependent interaction $v_0{\bm \sigma}_1\cdot{\bm \sigma}_2{\bm \tau}_1\cdot{\bm \tau}_2$ with $v_0\approx 240$ MeVfm$^3$ \cite{ichimura} can
increase the excitation energies of the M1 states by 0.5 MeV in rRPA. 
The summed values of $B({\rm M}1)\uparrow$ in rRPA and SCRPA are 5.63 $\mu_0^2$ and 4.53 $\mu_0^2$, respectively, which largely exceed the summed value $1.4~\mu_0^2$ of the observed two M1 states.
This overestimation of theoretical transition strengths is common for spin-isospin modes and is known as a quenching problem \cite{ichimura}.
In the following we point out that the large transition strength in rRPA and SCRPA is not inconsistent with the energy-weighted sum rule (EWSR).
We evaluate the EWSR value using
\begin{eqnarray} 
\sum_\mu |\langle \Phi_\mu|\hat{Q}_z|\Phi_0\rangle|^2&=&\frac{1}{2}\langle\Phi_0|[\hat{Q}_z,[H,\hat{Q}_z]]|\Phi_0\rangle
\nonumber \\
&=&\frac{1}{2}\langle\Phi_0|[\hat{Q}_z,[V,\hat{Q}_z]]|\Phi_0\rangle.
\label{ewsr}
\end{eqnarray}
Here $V$ is a two-body interaction and $\hat{Q}_z=\sum_i\sigma_i^z\tau_i^z$, which is used to calculate the M1 transition strength.
The EWSR value is determined by the spin-dependent part of the Hamiltonian $H$ and thus depends on the interaction $V$ \cite{adachi}.
As the spin-dependent interaction we first consider the spin-orbit force in Skyrme III
\begin{eqnarray}
V=i W_0({\bm \sigma}_1+{\bm \sigma}_2)\cdot {\bm k}'\times \delta^3({\bm r}_1-{\bm r}_2){\bm k},
\label{ls}
\end{eqnarray}
where ${\bm k}=(\nabla_1-\nabla_2)/2i$ acts on the right and ${\bm k}'=-(\nabla_1-\nabla_2)/2i$ on the left, and $W_0=120~ $MeVfm$^5$. 
The double commutator in Eq. (\ref{ewsr}) gives a two-body operator. 
To be consistent with the approximations for the ground state used in rRPA and SCRPA, 
we take only the uncorrelated parts of the two-body density matrix given by $n_\alpha$ in rRPA and both $n_\alpha$ and $C_2$ in SCRPA.
It turns out that the transition strengths in rRPA and in SCRPA exhaust 57 \% and 65\% of the EWSR values, respectively.
Thus the large transition strengths in rRPA and SCRPA are not inconsistent with EWSR associated with the spin-orbit force. 
To fulfill EWSR more precisely, we need to make the residual interaction to be consistent with the force used for the calculation of single-particle states
and the treatment of continuum states also affects the EWSR value. We found that when the box size used for the continuum states is decreased to 10 fm, the fulfillment of EWSR
is improved to 62 \% in rRPA. 
\begin{figure} 
\begin{center} 
\includegraphics[height=6cm]{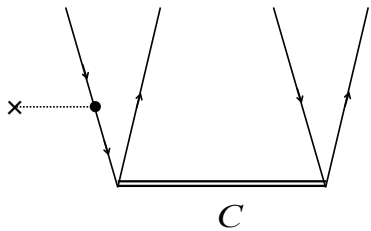}
\end{center}
\caption{Coupling of the h-h amplitude to the 2p-2h amplitude (4 open ended vertical lines) through $C_{\rm pp'hh'}$. 
The horizontal line indicates $C_{\rm pp'hh'}$
and the vertical lines with arrows either a hole state or a particle state. The dotted line with a cross at the left end depicts the
external field and the dot the h-h amplitude.} 
\label{diag} 
\end{figure} 

Now we discuss the results in ERPA.
In ERPA the most collective state appears at 8.7 MeV with 1.83 $\mu_0^2$. The excitation energy is 1.6 MeV lower than the experimental value 10.32 MeV. As mentioned above,
the excitation energies of M1 states depend on spin (and isospin) properties of the interactions used, which are not properly adjusted in this work.
This state has nature of the single-particle transitions similar to the collective states in rRPA and SCRPA.
The neutron and proton $1f_{7/2}\rightarrow 1f_{5/2}$ components have 54.2 \% and $10.2~\%$ contributions to the normalization, respectively.
The 2p--2h configurations $(\pi 2s_{1/2})^{-1}(\pi 1d_{3/2})^{-1}\pi 1f_{7/2}\pi 1f_{7/2}$ and $(\pi 2s_{1/2})^{-1}(\nu 2s_{1/2})^{-1}\pi 1f_{7/2}\nu 1f_{7/2}$ explain 6.2 \% of the normalization. 
The state at 8.6 MeV with 0.55 $\mu_0^2$ has also some single-particle nature: the neutron $1f_{7/2}\rightarrow 1f_{5/2}$ component has 
9.2 \% contribution to the normalization and $(\pi 2s_{1/2})^{-1}(\nu 2s_{1/2})^{-1}\pi 1f_{7/2}\nu 1f_{7/2}$ is the largest 2p--2h configuration which  
has 37.9 \% contribution to the normalization.
All other states have small single-particle components and mostly consist of  
$(\pi 1d_{3/2})^{-1}(\nu 1d_{3/2})^{-1}\pi 1f_{7/2}\nu 1f_{7/2}$ and $(\pi 1d_{3/2})^{-1}(\nu 2s_{1/2})^{-1}\pi 1f_{7/2}\nu 1f_{72}$ configurations.
Although these 2p--2h states have small single-particle transition components, 
the carry rather large M1 strengths. This is because the 2p--2h amplitudes can have the one-body transition amplitudes of the $ {x}^\mu_{\alpha\alpha}$ type
as graphically shown in Fig. \ref{diag}, where the horizontal line depicts $C_{\rm pp'hh'}$, the vertical lines with arrows either a particle or a hole state,
the four open ended vertical lines the 2p--2h amplitude, the dotted line with a cross at the left end the
external field and the dot the h-h amplitude. These $ {x}^\mu_{\alpha\alpha}$ type transition amplitudes are not included in rRPA and SCRPA.
These processes may be called the direct 2p--2h response \cite{nishi}.
It has been pointed out that these processes play an important role in the fragmentation of quadrupole strength  
in $^{16}$O \cite{toh07} and $^{40}$Ca \cite{toh17}.  Figure \ref{cam1} shows that the direct 2p--2h response also plays an important role in the fragmentation of M1 strength.
The most collective state in ERPA has by a factor of 1.6 larger transition strength than the experiment \cite{gross}. The results of shell-model calculations
for spin-isospin modes such as Gamow-Teller (GT) transitions require similar reduction factor to fit experimental data \cite{wakasa}.

The summed value of $B(M1)\uparrow$ in ERPA is 12.6 $\mu_0^2$ which is by a factor of about 2 larger than the rRPA value. We point out that
this large summed value in ERPA is in accordance with the results of the perturbative calculations \cite{adachi} where realistic tensor forces were used.
In the following we show that the transition strength in ERPA is not inconsistent with EWSR. The spin-orbit force cannot explain the large transition strength in ERPA.
We consider the other spin-dependent part of the Skyrme III force:
\begin{eqnarray}
V=t_0(1+x_0 P^\sigma)\delta^3({\bm r}_1-{\bm r}_2),
\label{t0x0}
\end{eqnarray}
where $P^\sigma$ is the spin-exchange operator, $t_0=-1128.75 $MeVfm$^3$ and $x_0=0.45$. The parameter $x_0$ was determined to reproduce empirical symmetry energy \cite{skIII}.
The ${\bm \sigma}_1\cdot {\bm \sigma}_2$ term in Eq. (\ref{t0x0}) contributes to
the right-hand side of Eq. (\ref{ewsr}) though it only induces small p--h correlations in rRPA and SCRPA for the isovector dipole mode.
In the case of ${\bm \sigma}_1\cdot {\bm \sigma}_2$
the EWSR value is determined solely by $C_2$:
the uncorrelated parts of the two-body density matrix have no contributions.
In the calculation of the EWSR value we use the reduction factor $f=0.66$. 
We find that the ERPA strength fulfills 75 \% 
of the EWSR value including the contribution of the spin-orbit force.
Here we emphasize that the large transition strength in ERPA is within the limit of the EWSR value.

\begin{figure} 
\begin{center} 
\includegraphics[height=6cm]{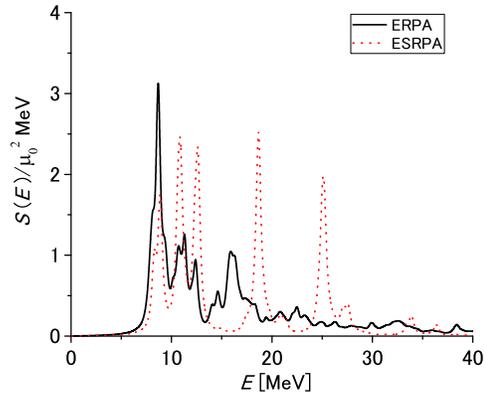}
\end{center}
\caption{Strength functions calculated in ERPA (solid line) and ERPA0 (dotted line) for the magnetic dipole excitation in $^{40}$Ca. The interaction among 2p--2h
configurations is neglected in ERPA0.
The distributions are smoothed with an artificial width $\Gamma=0.5$ MeV.} 
\label{cam3} 
\end{figure} 

Finally we point out the importance of correlations among 2p--2h configurations which are neglected in ESRPA \cite{srpa} and the response function formalism \cite{taka1,taka2}.
The dotted line in Fig. \ref{cam3} shows the results of a calculation where the correlations among 2p--2h configurations are switched off: This calculation is referred to as ERPA0. 
The peaks above 10 MeV in ERPA0 correspond to the excitation energies of the unperturbed 2p--2h states.
For example the unperturbed excitation energy of $(\pi 2_{1/2})^{-1}(\nu 2_{1/2})^{-1}\pi 1f_{7/2}\nu 1f_{7/2}$ is 10.6 MeV (see Table \ref{tab1}). These unperturbed 2p--2h configurations can have the M1 transition
strength through the process depicted in Fig. \ref{diag}.
Figure \ref{cam3} shows that the correlations among 2p--2h configurations drastically change the distribution of the M1 strengths and enhance the collectivity of the lowest states.

\vspace{0.5cm}
\section{Summary}
The effects of ground-state correlations on the magnetic dipole excitations were studied for $^{40}$Ca by using the extended
random phase approximation (ERPA) derived from the time-dependent density-matrix theory. Comparison with other extended RPA theories, rRPA, SCRPA and ESRPA was also made.
It was shown that the fractional occupation of the $1f_{7/2}$ states opens the single-particle transitions $1f_{7/2}\rightarrow 1f_{5/2}$, which are included in rRPA and SCRPA.
It was found that the M1 transition strengths in ERPA are strongly fragmented. It was pointed out that the diagonal part
of the one-body amplitudes plays an important role in increasing the transition strength in ERPA.
The effects of the correlations among the two-body amplitudes which are missing in ESRPA were studied and it was found that they play an important role in redistributing the M1 strength
and enhancing the collectivity of the lowest states.
Although the M1 transition strengths in rRPA, SCRPA and ERPA largely exceeded the experimental value, it was shown that they are not 
inconsistent with the energy-weighted sum-rule value obtained from the 
interaction used.
Since our main purpose was to compare extended RPA theories, interactions used were simple and self-consistency was not fully respected. Therefore, 
our analysis remained qualitative one.
Obviously realistic interactions with appropriate spin-isospin properties are needed to obtain more quantitative results.

\end{document}